\documentclass[sigconf]{acmart}
\AtBeginDocument{%
  \providecommand\BibTeX{{%
    \normalfont B\kern-0.5em{\scshape i\kern-0.25em b}\kern-0.8em\TeX}}}

\copyrightyear{2023} 
\acmYear{2023} 
\setcopyright{acmlicensed}\acmConference[WWW '23 Companion]{Companion Proceedings of the ACM Web Conference 2023}{April 30-May 4, 2023}{Austin, TX, USA}
\acmBooktitle{Companion Proceedings of the ACM Web Conference 2023 (WWW '23 Companion), April 30-May 4, 2023, Austin, TX, USA}
\acmPrice{15.00}
\acmDOI{10.1145/3543873.3587608}
\acmISBN{978-1-4503-9419-2/23/04}

\acmConference[Sci-K '23]{3rd International Workshop on Scientific Knowledge: Representation, Discovery, and Assessment (Spring 2023) co-located with The Web Conference 2023}{Austin, TX}

\usepackage{cleveref}
\usepackage{tabularx}

\begin{document}

\title{Assessing Scientific Contributions in Data Sharing Spaces}


\author{Kacy Adams}
\email{adamks4@rpi.edu}
\orcid{0000-0003-2873-8616}
\affiliation{%
  \institution{Rensselaer Polytechnic Institute}
  \streetaddress{110 8th St}
  \city{Troy}
  \state{NY}
  \country{USA}
}

\author{Fernando Spadea}
\email{spadef@rpi.edu}
\orcid{0009-0006-4278-3666}
\affiliation{%
  \institution{Rensselaer Polytechnic Institute}
  \streetaddress{110 8th St}
  \city{Troy}
  \state{NY}
  \country{USA}
}
\author{Conor Flynn}
\email{flynnc@rpi.edu}
\orcid{0009-0006-4537-2838}
\affiliation{%
  \institution{Rensselaer Polytechnic Institute}
  \streetaddress{110 8th St}
  \city{Troy}
  \state{NY}
  \country{USA}
}

\author{Oshani Seneviratne}
\email{senevo@rpi.edu}
\orcid{0000-0001-8518-917X}
\affiliation{%
  \institution{Rensselaer Polytechnic Institute}
  \streetaddress{110 8th St}
  \city{Troy}
  \state{NY}
  \country{USA}
}


\begin{abstract}  
In the present academic landscape, the process of collecting data is slow, and the lax infrastructures for data collaborations lead to significant delays in coming up with and disseminating conclusive findings. Therefore, there is an increasing need for a secure, scalable, and trustworthy data-sharing ecosystem that promotes and rewards collaborative data-sharing efforts among researchers, and a robust incentive mechanism is required to achieve this objective. Reputation-based incentives, such as the h-index, have historically played a pivotal role in the academic community. However, the h-index suffers from several limitations.
This paper introduces the SCIENCE-index, a blockchain-based metric measuring a researcher's scientific contributions. Utilizing the Microsoft Academic Graph and machine learning techniques, the SCIENCE-index predicts the progress made by a researcher over their career and provides a soft incentive for sharing their datasets with peer researchers. 
To incentivize researchers to share their data, the SCIENCE-index is augmented to include a data-sharing parameter. DataCite, a database of openly available datasets, proxies this parameter, which is further enhanced by including a researcher's data-sharing activity.
Our model is evaluated by comparing the distribution of its output for geographically diverse researchers to that of the h-index. We observe that it results in a much more even spread of evaluations. The SCIENCE-index is a crucial component in constructing a decentralized protocol that promotes trust-based data sharing, addressing the current inequity in dataset sharing.
The work outlined in this paper provides the foundation for assessing scientific contributions in future data-sharing spaces powered by decentralized applications.
\end{abstract}

\ccsdesc[300]{Social and professional topics~Industry statistics}
\ccsdesc[300]{General and reference~Metrics}
\ccsdesc[300]{Information systems~Social recommendation}
\ccsdesc[300]{Information systems~Learning to rank}

\keywords{Incentive Mechanisms, Author-level Metrics, Dataset Sharing, Peer-to-peer Sharing, Blockchain, Smart Contracts}

\maketitle

\section{Introduction}

There is a rising need for a secure, scalable, and trustless data-sharing ecosystem that recommends, incentivizes, and rewards collaborative data-sharing efforts between researchers in many scientific disciplines.
Such a protocol would require a robust incentive mechanism. 
In the academic space, reputation-based incentives rule, and since 2006, h-index~\cite{kelly2006h} has reigned superior. It is a widely used bibliometric indicator that measures a scientist's publications' productivity and citation impact. The h-index, however, has several flaws.
One of the main shortcomings of the h-index is that it does not account for the quality or impact of individual publications but considers all publications equally. Additionally, the h-index tends to favor established researchers with a long publication history, as it considers the total number of publications. The h-index may be subject to manipulation by self-citations or citation cartels, which can inflate an author's score. Therefore, it is important to use the h-index in conjunction with other metrics and to interpret it with caution.

If all research is to be fair and incentivized, researchers must mend these discrepancies as their reputations define their careers. We present a new reputation-based metric called the SCIENCE-index to incentivize and reward data aggregation and sharing.

We utilize the Microsoft Academic Graph (MAG)~\cite{sinha2015an} to train an AI model to predict the researcher's progress over their career. We persist this model via smart contract on the Ethereum blockchain~\cite{buterin2016ethereum} and allow researchers to look up their and others' SCIENCE-indexes via web identifiers such as the Semantic Scholar ID.
A blockchain-based mechanism provides several advantages, including increased security and transparency, as well as the ability to incentivize data sharing through the use of smart contracts.

\subsection{The Need for Data Sharing}

Data is a significant part of modern evidence-based scientific research. Many studies rely heavily on collecting large amounts of data ranging from studies on human behaviors to machine learning. Collecting this data can be painstaking and time-consuming, taking up to seventeen years from bench to bedside in certain biomedical research studies~\cite{morris2011answer}. Even in computer or information science research, a significant effort goes into collecting data. For example, several web science researchers 2012 conducted a study on buying unlicensed slimming drugs online, which required ethnographic data collection. They had to manually copy and paste parts of their data from the sites they scoured and held interviews with several stakeholders in UK regulatory agencies~\cite{drugs}. 

Despite data collection being very important, researchers are typically not incentivized to share their data. In fact, there are many reasons not to share data. Researchers face many challenges when it comes to intellectual property and confidentiality~\cite{meadows-to-share-or-not-share}. Others from less-endowed institutions may fear their work being scooped up by more prestigious institutions~\cite{bezuidenhout2018hidden} or fear that others may use it to their advantage~\cite{fernandez2010barriers}. 

On top of this, existing academic incentives that reach further than a citation are scarce. One incentive, the use of open data badges\footnote{\url{https://osf.io/tvyxz/wiki/home}}, has been tested in health and medical research~\cite{kidwell2016badges}, yet studies are unsure of their effectiveness~\cite{rowhani2017incentives}.  
Better incentives are necessary as data sharing is essential in modern academic research.
In the case of protecting research participants, data sharing is an ethical necessity to protect human lives. In January 2016, one participant died, and four others were injured due to the first human testing of a fatty acid inhibitor \cite{datasharing}. With human lives at stake, data from studies like these must be made available.

In this work, we address the aforementioned inequity of dataset sharing. To incentivize researchers to share their data, we must build their reputation to include their data-sharing activity and the bibliometric reputation available through indexes such as the h-index. We augment our initial SCIENCE-index to include a data-sharing parameter. We proxy this via DataCite~\cite{brase2009datacite}, a database of openly available datasets, and widen our MAG dataset by including the number of times a researcher has shared their data.
Such a metric would be pivotal in building a decentralized protocol that allows data sharing by adding trust between individuals who have not worked together prior to the collaboration event.

\subsection{More Than The h-index}

The h-index was created to measure the impact of a researcher's work. It represents the maximum number of ``h'' papers published by a researcher with at least ``h'' citations. This used to be a fairly accurate reflection of researchers' past impact, at least when tested against the reviews of the \emph{Bochringer Ingelheim Fonds} organization \cite{hindex}. However, the h-index's reliability has greatly diminished in recent years and no longer represents the scientific reputations of researchers \cite{hbad}. 

Further, we express our concerns about the unfair playing field of scientific research in underdeveloped countries~\cite{unfairplayingfield, developing}. We believe that more robust reputation-based metrics would help to level the ``playing field'' among researchers with fewer resources compared to well-endowed researchers.

Data sharing is a massive service to the research community largely unconsidered by the h-index. A system is needed to measure an individual's contributions to science beyond that of publications. Then, researchers will be properly credited for their work and, thus, incentivized to continue to collect and share invaluable data.

\subsection{A Data Sharing Space}

The work outlined in this paper is part of a larger plan to create a data-sharing environment via blockchain. This environment would allow researchers to share data while being rewarded. We believe blockchain is an appropriate technology to use here because it is an effective ledger for keeping track of data sharing in a transparent and accountable manner. Its decentralized and public nature also makes it more transparent for the researchers who share and use data on it. To incentivize participation and the sharing of good quality data, we propose a new index, i.e., the SCIENCE-index, for assessing the impact of researchers' contributions with a model that will be hosted and persisted in the same decentralized manner.

We specifically target fields where research requires and produces large amounts of data. This is because the number of opportunities to share and use data varies greatly between fields, so there is no one-size-fits-all measure of one's contributions to data sharing. However, by specifically considering data-heavy fields, like, for example, biomedical research, we can create an accurate metric for those fields without downplaying the significance of work in other fields with fewer datasets.

We present the SCIENCE-index in \Cref{sec:SCIdex}, detailing a robust linear model for rating researchers in \Cref{sec:Model1}. In \Cref{sec:infra}, we display a decentralized infrastructure for persisting this model in a public fashion. Next, we discuss our prior work in the decentralized data sharing space in \Cref{sec:sharingscience} and augment our linear model to incentivize data sharing. We present results from the model and its augmentation in \Cref{sec:results} and evaluate the efficacy of our model on geographically distributed researchers in \Cref{sec:eval}. Finally, we discuss our work and future work in \Cref{sec:disc} and present related work in \Cref{sec:related}.

\section{The SCIENCE-index}
\label{sec:SCIdex}

We present the SCIENCE-index, a self-sustaining metric for scientific reputation. The SCIENCE index encompasses an expressive, provenance-centric approach, and it is a recursive acronym for \underline{S}CIENCE, \underline{C}apability-based, \underline{I}ntention-centric, \underline{E}xperiment-oriented, \underline{N}etworked, \underline{C}ollaborative, \underline{E}xpression.

We bootstrap the SCIENCE-index via 21 million data points from the MAG that overlaps with entries from the data-sharing website DataCite.
First, using a robust multiple linear regression across several academic career statistics, we predict a researcher's h-index and compare this to their actual h-index. 
This difference is then normalized to a scale from zero to ten. Five means expected, under five is below average, and over five is above average. 
We persist the model via smart contract on a public blockchain, allowing the model to exist publicly and continue to scale and update as researchers use it. We detail the model and its infrastructure below.

\subsection{Data Aggregation}

We utilize the MAG to aggregate a dataset of 21,660,755 authors, each with their corresponding publication count, citation count, h-index, and career length. We select and use publication count and citation count as they are straightforward indicators of productivity and are provided as part of the ``authors'' table of the MAG. With some simple calculations, we can build paper lists for each of our authors via the ``papers'' table of the MAG, and with this, we can assume career length as the years between their oldest paper and their newest paper and calculate their h-index. Although within the MAG and the academic research space, there are many more abstract parameters to use, these four were the most accessible via the web through tools such as the Semantic Scholar API\footnote{\url{https://www.semanticscholar.org}}. This accessibility is an important piece of our goal to persist our metric in a decentralized manner.

\subsection{The Model}
\label{sec:Model1}

The model of the SCIENCE-index takes in four different inputs: career length, paper count, citation count, and h-index.
\[ \alpha_1 \textit{= Career Length} \]
\[ \alpha_2 \textit{= Paper Count} \]
\[ \alpha_3 \textit{= Citation Count} \]
\[ \alpha_4 \textit{= h-index} \]
We calculate the predicted h-index value ($\beta$) from these parameters and compare it to the actual h-index ($\alpha_4$), extracting the SCIENCE-index from this difference. Since we have a narrow dataset, we use Multi-Linear Regression (MLR) to find our predicted h-index. After training the MLR on our 21 million data points, we derive the following equation
\begin{align*}
&\beta &&=\omega_0+\omega_1\alpha_1+\omega_2\alpha_2+\omega_3\alpha_3\\
& &&=1.71933+0.06902\alpha_1+0.10867\alpha_2+0.00304\alpha_3
\end{align*}
Finally, we scale for outliers that can occur anywhere $\beta>60$. This threshold, i.e., \textbf{60}, is based on the notion that ``an h index of \textbf{60} after 20 years, or 90 after 30 years, characterizes truly unique individuals" as stated by Hirsch~\cite{pnas}. Therefore, we give these researchers a bias to their SCIENCE-index calculation such that it will give them a higher score. If $\beta>60$, we apply the given function to scale it to an appropriate value.

\begin{align*}
&\text{If }\beta>60\text{:}&&
\beta =\frac{\beta}{0.571+(0.007*\beta)}
\end{align*}

Using these weights and our approximation for $\beta$, we can then calculate the difference ($\delta$) between the predicted and the actual h-index.
After finding $\delta$, we normalize it according to the entire dataset to find $\epsilon$, our calculated performance factor comparable to any other data point's $\epsilon$ value. We then logarithmically regress the scaled delta to fit on a scale of one to ten for readability and easy comparison.

\[\delta =\alpha_4-\beta\]
\[\epsilon =\frac{\delta-\overline{\delta}}{\sigma_\delta}\]
\[\phi =\frac{10}{1+e^{-\epsilon}}\]

This calculated value of $\phi$ is the outputted SCIENCE-index. Any value of $\phi$ below $5$ is deemed a below-average academic contribution by the researcher, and any value of $\phi$ above $5$ signifies above-average contributions.
 
\subsection{Infrastructure}
\label{sec:infra}

\begin{figure}[htbp!]
    \centering
    \includegraphics[width=\columnwidth]{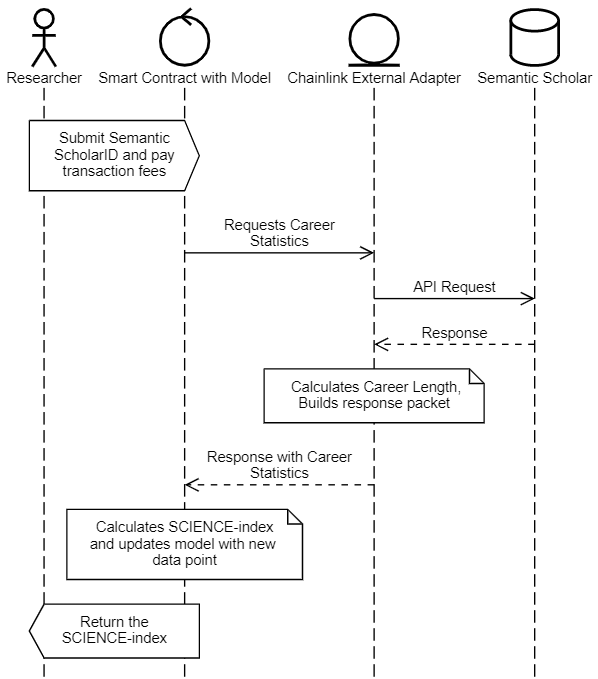}
    \caption{High-level Sequence Diagram of Our Linear Model/Smart Contract/Oracle Design Pattern}
    \label{fig:SCIDEXsequence}
\end{figure}

The SCIENCE-index lives as weights in a smart contract, making it publicly accessible and completely transparent. 
Upon call, a researcher provides their Semantic Scholar identifier, and the smart contract requests a Chainlink oracle~\cite{chainlink}, which requests the Semantic Scholar API to get the requesting researcher's statistics. 
Using these statistics, the smart contract adds the new data point to the model by updating the weights and then calculates and returns the researcher's SCIENCE-index. This sequence is shown in \Cref{fig:SCIDEXsequence}.

Chainlink is an industry-standard service for building blockchain oracles. Oracles allow smart contracts that live on the immutable blockchain to access the outside, mutable internet. 
We utilize a Chainlink external adapter to build a custom job that, when called, calculates and returns the career statistics used in our model.

\section{Sharing Science}
\label{sec:sharingscience}

We narrow the scope of our incentive mechanism towards a hypothetical data-sharing ecosystem.
The SCIENCE-index is part of the Sharing Science Ontology (SSO), a semantic model for a decentralized academic data-sharing application.

\subsection{The Sharing Science Ontology (SSO)}

We have discussed the need for a data-sharing ecosystem in a secure and scalable manner, and blockchain provides us with the peer-to-peer platform to do so. 
The SSO describes a decentralized protocol to handle and incentivize peer-to-peer academic data sharing in a distributed environment.

The SSO handles peer-to-peer sharing events via what is called collaboration events.
Collaboration events are between two parties and begin with a data request by the data seeker and end with a citation of the data sharer by the data seeker.
Researchers are rewarded via incentive mechanisms for their honest and fair completion of collaboration events.
The SSO is publicly available at a persistent URL at \url{https://github.com/sharing-science/sharing-science-ontology}. 

\subsection{The SCIENCE-index Augmented}
\label{sec:Model2}

In an attempt to incentivize the sharing of data, we introduce a new parameter to our SCIENCE-index model. We include the number of times a researcher has shared a dataset. In our discussed protocol, this would represent the number of collaboration events a researcher has participated in. 
We bootstrap the model again with approximately 3000 data points from the MAG. Using the DataCite API~\cite{brase2009datacite}, we can count the times a researcher has published a publicly available dataset and establish this as a proxy for our collaboration events. Our parameters now include the following:
\[ \alpha_1 \textit{= Career Length} \]
\[ \alpha_2 \textit{= Paper Count} \]
\[ \alpha_3 \textit{= Citation Count} \]
\[ \alpha_4 \textit{= Data Share Count}^2 \]
\[ \alpha_5 \textit{= h-index} \]
We again regress on the h-index to predict a researcher's h-index and scale the difference. To further incentivize data sharing, we weigh the number of data shares by a power of two, and this gives enough weight to data sharing that researchers can effectively increase their SCIENCE-index through data-sharing activities.

\section{Results}
\label{sec:results}
With train and test sets from our initial data sets, we can visualize the results of our two proposed models.

\subsection{The SCIENCE-index}
\begin{figure}[htbp!]
    \centering
    \includegraphics[width=\columnwidth]{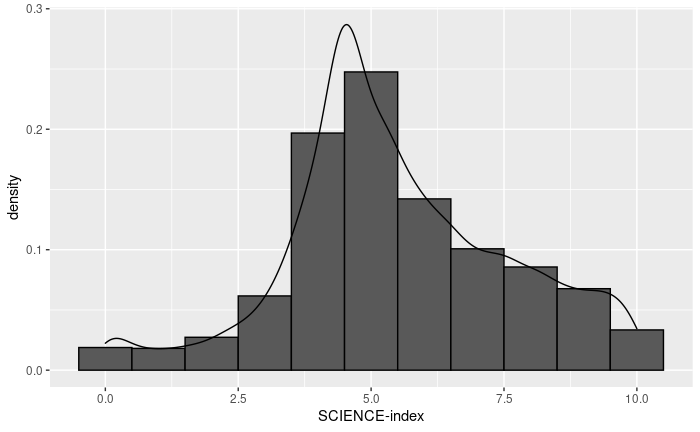}
    \caption{Density Plot of the SCIENCE-Index Described In \Cref{sec:Model1}}
    \label{fig:SCIDEXdensity}
\end{figure}

Without the data sharing parameter, our initial SCIENCE-index presents a distribution as seen in \Cref{fig:SCIDEXdensity}.

\begin{figure}[htbp!]
    \centering
    \includegraphics[width=\columnwidth]{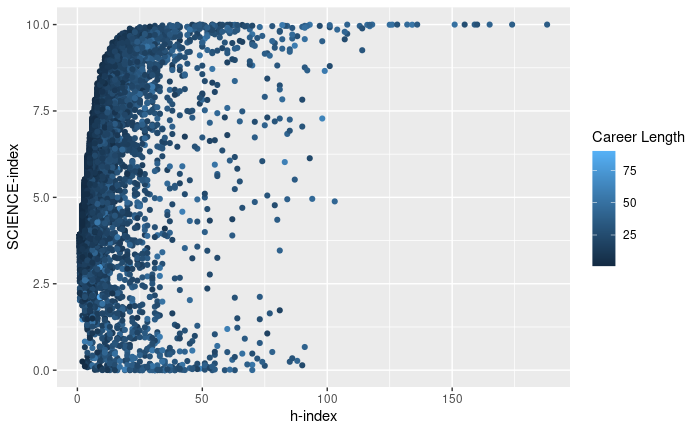}
    \caption{h-index Against the SCIENCE-Index Characterized by Career Length \Cref{sec:Model1}}
    \label{fig:SCIDEXvsH}
\end{figure}

We can see that the model is conservative and leans forward after its density peaks just before 5. We further visualize the model in \Cref{fig:SCIDEXvsH}.

We see that career length is correlated with the h-index, as the shade of blue gets lighter from left to right. However, career length is not correlated with the SCIENCE-index, which allows us to compare researchers of any age. 

\subsection{The SCIENCE-index with Data Sharing Data}

\begin{figure}[htbp!]
    \centering
    \includegraphics[width=\columnwidth]{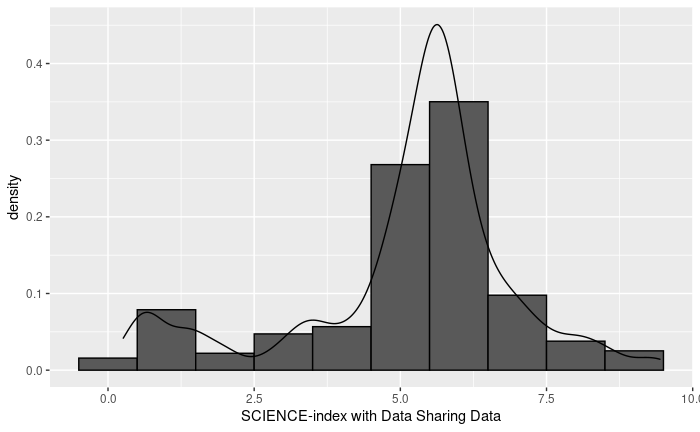}
    \caption{Density Plot of the Augmented SCIENCE-Index Described In \Cref{sec:Model2}}
    \label{fig:SCIDATAdensity}
\end{figure}

Similar to before, once augmented by the data sharing data, we have a forward-leaning density of the metric shown in \Cref{fig:SCIDATAdensity}

\begin{figure}[htbp!]
    \centering
    \includegraphics[width=\columnwidth]{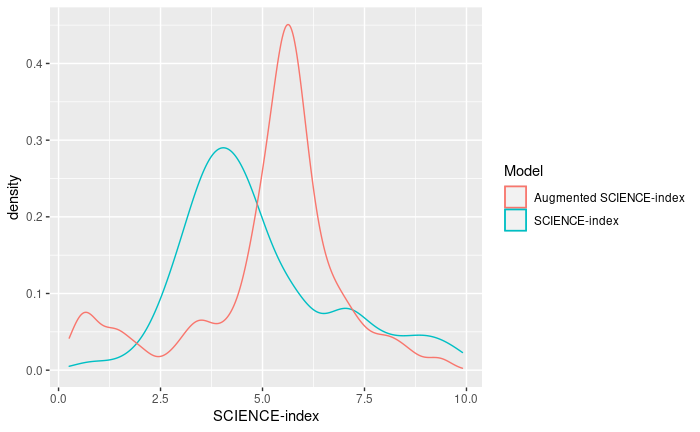}
    \caption{Density Plots of Both Models Overlayed To Show the Effect of the Introduction of Data Sharing Data}
    \label{fig:SCIDATAcomp}
\end{figure}

We can compare the original model to the now augmented model using our initial SCIENCE-index with our data-sharing proxy dataset. In \Cref{fig:SCIDATAcomp}, we compare the density of the two models. The augmented SCIENCE-index has been shifted forward as each member has shared data.

The data sharing dataset has an average data share ``count'' of $6.6$, giving us an average positive shift in each researcher's SCIENCE-index of $0.27$.

\section{Evaluation}
\label{sec:eval}

In the spirit of addressing inequality using the SCIENCE-index, we tailor our evaluation toward comparing geographically distributed researchers. We argue that researchers from less developed countries with fewer resources face larger challenges in building their academic reputations. We have compiled a brief dataset of researchers with half the dataset affiliated with universities located in the ``global south,'' i.e., resource-poor institutions, such as Rhodes University (\url{https://www.ru.ac.za}), University of Sao Paulo (\url{https://www.fearp.usp.br}) and the other half located in the ``global north,'' i.e., resource-rich institutions, such as Stanford University, (\url{https://www.stanford.edu}), Mcgill University (\url{https://www.mcgill.ca}),  University of North Carolina (\url{https://www.unc.edu}), and Grenoble Alpes University (\url{https://www.univ-grenoble-alpes.fr}).

\begin{figure}[htbp!]
    \centering
    \includegraphics[width=\columnwidth]{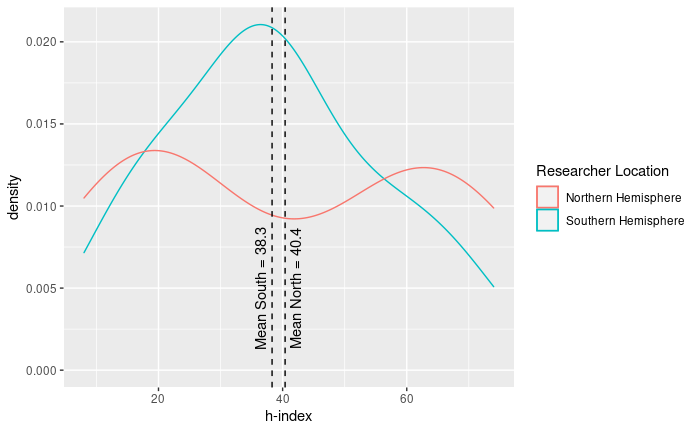}
    \caption{Density Plot of the h-indexes of Our Hemisphere-Separated ``Global South'' and ``Global North'' Dataset}
    \label{fig:EvalH}
\end{figure}

In \Cref{fig:EvalH}, we show the difference in the h-index between the two groups of researchers with a density plot. The mean of the northern-located researchers is an average of 2 points greater than the southern-located researchers.

\begin{figure}[htbp!]
    \centering
    \includegraphics[width=\columnwidth]{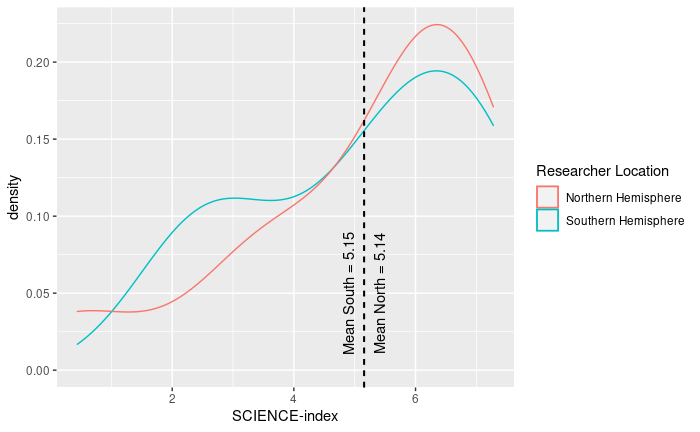}
    \caption{Density Plot of the SCIENCE-Index of Our Hemisphere-Separated Dataset}
    \label{fig:EvalSCI}
\end{figure}

We then run this group of researchers through our SCIENCE-index model trained on our original dataset. We show the results of this in \Cref{fig:EvalSCI}. The mean of the SCIENCE-index of each group converges to ~$5.1$. This shows the ability of the SCIENCE-index to level the ``playing field'' and look objectively at a researcher's career progress.

\section{Discussion}
\label{sec:disc}

The SCIENCE-index aims to be a valuable tool to credit researchers fairly for their contributions. It allows us to compare researchers at different points in their careers and levels of discrepancies in career statistics. Such a model has an incredibly diverse set of applications, as we can widen the training dataset as we desire. Different parameters can be used and weighted to augment the h-index in favor of specific activities, such as data sharing, as we've discussed, or other statistics, such as conference reviewing or institutional affiliations. This extensibility opens the door to incentive mechanisms in all aspects of academic research.
More importantly, the SCIENCE-index predicts the future progress of a researcher based on their past contributions to science. This is important because it encourages researchers to continue contributing significantly to science and rewards those with a consistent track record.

\subsection{Different Models}

We consider future work examining different learning models to back the SCIENCE-index. This paper explores using multi-linear regression, but many different models could be trained on the data, including logistic regression, support vectors, or decision tree clustering. Yet, we initially aimed to avoid a black box model, as researchers are less likely to accept a metric they cannot interpret.
We also note that there is no one size fits all solution for scientific contributions. For example, it is well known that the citation culture varies between areas, e.g., biology is different from computer science, so the model outlined in \Cref{sec:Model1} may need tweaks to cater to the specific needs of the different scientific communities.

\subsection{Bootstrapping the Models}

A large amount of data is required to bootstrap any model. For the SCIENCE-index, we use a subset of the MAG. The MAG, however, has been deprecated, and for future training of models, we need a current, accurate date. Examining other parameters, such as data sharing, we must look to other sources. We also use DataCite~\cite{brase2009datacite}, but future iterations require more robust and accurate aggregation of data-sharing statistics. Several other data-sharing resources covered in \Cref{sec:related} could be used to aggregate prior data-sharing activity.

We acknowledge that adding a data component to the model in evaluating the SCIENCE-index can introduce some flaws as it may be easier to publish data than a paper, which could lead to artificially inflating one's SCIENCE-index for a useless dataset. However, the broader framework would allow the data reusers to review the dataset they have access to, and future iterations of the model would incorporate the citations to the datasets directly.

Another issue we foresee is the need for a canonical identifier. In our work, we utilized the Semantic Scholar ID. However, author disambiguation is a major issue in academic publishing, which will only worsen once datasets are factored in.
We plan to utilize robust entity resolution mechanisms that will leverage a variety of identifiers, including Orcid~\cite{haak2012orcid} and decentralized identity mechanisms~\cite{reed2020decentralized} championed by standards organizations such as the W3C.

\subsection{Blockchain Usage}

Our goal of persisting a reputation-based metric in a decentralized manner has come with challenges. Public blockchains present invaluable peer-to-peer networks that allow transparency and scalability, and they also present limitations on computability and cost. Our model must be computationally lightweight not to incur incredibly high transaction costs when computing on a public virtual machine but also be robust enough to rate researchers accurately. We must also consider the oracle problem~\cite{oracle2022}, which presents difficulties accessing mutable data from an immutable state machine. Our use of Chainlink oracles described in \Cref{sec:infra} is our first take at tackling this problem.
There are also questions about who takes on the burden of paying for the transaction fees. When a dataset is being requested for a collaboration event, the benefiting party is the dataset requester, and it seems only fair that they should pay the fees required. However, the subsequent citations benefit the original dataset sharer more than the requester unless there is a punitive mechanism for failing to provide proper data citations. Therefore, robust tokenomics should be defined in future work to benefit all parties involved.

\section{Related Work}
\label{sec:related}

We describe two main themes throughout our work. The former seeks to improve equity and fairness in author-level metrics, while the latter seeks to encourage and incentivize open academic data sharing. 
We explore related work among both of these.

\subsection{Improving Reputation-based Metrics}

The SCIENCE-index is not competing with other researcher metrics but rather looking to host a metric in a public data-sharing environment on a blockchain. Other research in this field is not a competitor as it is an opportunity to improve how the proposed framework can incentivize researchers to share their work.

The g-index takes the maximum number of ``g'' papers that collectively have $g^2$ citations \cite{gindex}. This makes the g-index more sensitive to impactful papers while avoiding letting insignificant papers have too much influence. The g-index would always be at least the value of the h-index, but the g-index adds the extra push from more important papers. Some argue that the two indexes do not replace each other but rather complement each other, where the h-index favors big paper producers and the g-index favors selective researchers~\cite{costas_bordons_2008}. Others, such as Google Scholar\footnote{\url{https://scholar.google.com}}, take a more straightforward approach with the i10-index~\cite{connor2011google}. This is simply the number of papers from a researcher with at least ten citations. Unlike the g-index, this is very simple to use and understand, but like the g-index, it has not risen to acceptance in the same way as the h-index.

We look at another author-level metric that tries to identify meaningful citations. The paper claims that the importance of citations can be measured with three objective measurements and that by identifying these, we can properly distinguish types of citations and credit researchers accordingly~\cite{meaningfulcitations}. 
The significance of a researcher's citations is another important metric for us to consider when comparing researchers. Similarly, it is worthwhile to look into detecting the significance of cited data to the results of a paper. For example, data used to train a model successfully may be more significant than data used to test a model. Such a measurement could improve the SCIENCE-index greatly in the scope of data sharing.

Of course, there is significant work in creating better metrics for assessing research publications. Two major examples are the Field-Weighted Citation Impact (FWCI) and Relative Citation Ratio (RCR) \cite{purkayastha2019comparison}. The FWCI is an article-level metric that reflects the significance of a paper's citation count. It considers the publication type, year, and subject area to measure how a specific publication compares to others \cite{purkayastha2019comparison}. The RCR similarly divides the citation count of a publication by its expected value which is calculated with a quantile regression analysis of the citations of prior publications funded by the National Institutes of Health plotted against the field citation rate \cite{purkayastha2019comparison}. These article-level metrics can be collected to assess researchers, so they have significant promise and could be a useful source to improve the SCIENCE-index.

There is also other work in assessing researchers while considering their data contributions. One example is the data-index which takes into account data publications and citations \cite{hood2021data}. Other examples include the s-index~\cite{ko2013index} and Data Citation Index (DCI) \cite{olfson2017incentivizing, park2019research}. These are also promising projects in the same field as this paper, and each takes different approaches to reach a similar result. However, they are again not competitors but rather potential sources of improvement for the SCIENCE-index as an incentivizer. They could be useful in adjusting our model or even for bootstrapping purposes.

\subsection{Distributed Data Sharing}

Harvard's Dataverse project\footnote{\url{https://dataverse.org/about}}~\cite{king2007introduction} is a centralized approach to solving the problem of credited data sharing. It attempts to promote responsible data sharing by streamlining the process for researchers. 
It offers the option of archiving one's data in a ``collection'' and generating a unique academic citation with its own Document Object Identifier (DOI) that allows others to cite the data resource properly. 
The project aims to give researchers full control over their data.
Data sharers can make data completely open or require that each person that wishes to look at it must ask them for permission to access the data. 
Researchers are also given the tools to add metadata to their data so that other researchers can find it in search engines. This allows researchers to maintain control over how their data is distributed while benefiting from institutional backing, such as in DataVerse. 
Our discussed data sharing protocol in \Cref{sec:sharingscience} would share many traits of the Harvard Dataverse, such as role-based access control over their data. Further, the ability to create DOIs specifically for datasets makes it much more feasible to credit researchers for their data contributions. The Dataverse shows us the importance of providing proper infrastructure and tools for researchers to handle, annotate, and share data easily.

The Ocean Protocol~\cite{ocean} is a decentralized data economy. Similar to the Dataverse, it promotes responsible data sharing by attempting to create a hub for researchers' data where they can be rewarded for their contributions. With the Ocean Protocol, one can publish their data as an NFT and then sell access tokens for their data. By tracking the usages of the data on a blockchain, researchers can be credited for their data contributions by citation while fiscally rewarding authors. Ocean Protocol's decentralized approach to data sharing promotes responsible data sharing while strongly incentivizing it, but it still fails to be the data ecosystem we seek to create. As researchers must pay for data, many will not have the resources to ``buy'' datasets. 

\section{Conclusion}

Data sharing is a vital step towards more efficient and overall better research. 
SCIENCE-index addresses several flaws of the more major indexes, such as the inability to differentiate between highly cited but low-quality papers and low-cited but high-quality papers. The SCIENCE-index also considers the impact of data sharing, which is becoming increasingly important in scientific research.
Our framework is a decentralized, self-governed, peer-to-peer data-sharing protocol that would connect distributed researchers, decrease data reproduction, and increase research productivity.
We build an ecosystem that fosters and rewards collaborations. 
However, this framework would only survive and scale if researchers were properly incentivized to participate. Our SCIENCE-index attempts to improve on current reputation-based metrics of measuring researchers, specifically the h-index, by augmenting it with data-sharing capabilities. We predict a given researcher's h-index based on 21 million other researchers and compare this to their actual h-index. This comparison gives us insight into their career progress compared to their peers. We also find that our model has a much more even spread of evaluations than the h-index when applied to geographically diverse researchers indicating that we have created a fair metric.
We extend this to the scope of our data-sharing endeavors. By including data-sharing statistics as a parameter, we can reward researchers for their data sharing, thus incentivizing further data sharing. This incentive mechanism is necessary for distributed data sharing and encouraging more open science. 
This would increase the visibility of researchers who share their data and provide funding opportunities for those who share their data. While these incentives may not be perfect, they are a step in the right direction toward encouraging more data sharing in scientific research.
Our initial SCIENCE-index levels the playing field among researchers with various amounts of resources at various points in their careers. 
Finally, we assert that the SCIENCE-index and its underlying infrastructure open the door for further discussion regarding how we rate and incentivize researchers.

\medskip
\noindent\textbf{Resource Contributions:}
We contribute the SCIENCE-index as an open-source repository, including our code for data gathering, model training, and visualization. We also include the smart contract code, which persists in our model, and the rest of our decentralized application. 
Our research artifacts are shared under the Apache 2.0 license. We maintain open-source Github repositories for all our artifacts at \url{https://github.com/AI-and-Blockchain/F22_SCIENCE_Index}.



\bibliographystyle{ACM-Reference-Format}
\bibliography{references}


\begin{thebibliography}{32}


\ifx \showCODEN    \undefined \def \showCODEN     #1{\unskip}     \fi
\ifx \showDOI      \undefined \def \showDOI       #1{#1}\fi
\ifx \showISBNx    \undefined \def \showISBNx     #1{\unskip}     \fi
\ifx \showISBNxiii \undefined \def \showISBNxiii  #1{\unskip}     \fi
\ifx \showISSN     \undefined \def \showISSN      #1{\unskip}     \fi
\ifx \showLCCN     \undefined \def \showLCCN      #1{\unskip}     \fi
\ifx \shownote     \undefined \def \shownote      #1{#1}          \fi
\ifx \showarticletitle \undefined \def \showarticletitle #1{#1}   \fi
\ifx \showURL      \undefined \def \showURL       {\relax}        \fi
\providecommand\bibfield[2]{#2}
\providecommand\bibinfo[2]{#2}
\providecommand\natexlab[1]{#1}
\providecommand\showeprint[2][]{arXiv:#2}

\bibitem[{Alice Meadows}(2014)]%
        {meadows-to-share-or-not-share}
\bibfield{author}{\bibinfo{person}{{Alice Meadows}}.}
  \bibinfo{year}{2014}\natexlab{}.
\newblock \bibinfo{booktitle}{\emph{{To Share or not to Share? That is the
  (Research Data) Question...}}}
\newblock {The Scholarly Kitchen}.
\newblock
\urldef\tempurl%
\url{https://scholarlykitchen.sspnet.org/2014/11/11/to-share-or-not-to-share-that-is-the-research-data-question}
\showURL{%
Retrieved Jan 31, 2021 from \tempurl}


\bibitem[Amarante et~al\mbox{.}(2022)]%
        {developing}
\bibfield{author}{\bibinfo{person}{Verónica Amarante},
  \bibinfo{person}{Ronelle Burger}, \bibinfo{person}{Grieve Chelwa},
  \bibinfo{person}{John Cockburn}, \bibinfo{person}{Ana Kassouf},
  \bibinfo{person}{Andrew McKay}, {and} \bibinfo{person}{Julieta Zurbrigg}.}
  \bibinfo{year}{2022}\natexlab{}.
\newblock \showarticletitle{Underrepresentation of developing country
  researchers in development research}.
\newblock \bibinfo{journal}{\emph{Applied Economics Letters}}
  \bibinfo{volume}{29}, \bibinfo{number}{17} (\bibinfo{year}{2022}),
  \bibinfo{pages}{1659--1664}.
\newblock
\urldef\tempurl%
\url{https://doi.org/10.1080/13504851.2021.1965528}
\showDOI{\tempurl}
\showeprint{https://doi.org/10.1080/13504851.2021.1965528}


\bibitem[Bezuidenhout and Chakauya(2018)]%
        {bezuidenhout2018hidden}
\bibfield{author}{\bibinfo{person}{Louise Bezuidenhout} {and}
  \bibinfo{person}{Ereck Chakauya}.} \bibinfo{year}{2018}\natexlab{}.
\newblock \showarticletitle{Hidden concerns of sharing research data by
  low/middle-income country scientists}.
\newblock \bibinfo{journal}{\emph{Global Bioethics}} \bibinfo{volume}{29},
  \bibinfo{number}{1} (\bibinfo{year}{2018}), \bibinfo{pages}{39--54}.
\newblock


\bibitem[Bornmann and Daniel(2007)]%
        {hindex}
\bibfield{author}{\bibinfo{person}{Lutz Bornmann} {and}
  \bibinfo{person}{Hans-Dieter Daniel}.} \bibinfo{year}{2007}\natexlab{}.
\newblock \showarticletitle{What do we know about the h index?}
\newblock \bibinfo{journal}{\emph{Journal of the American Society for
  Information Science and technology}} \bibinfo{volume}{58},
  \bibinfo{number}{9} (\bibinfo{year}{2007}), \bibinfo{pages}{1381--1385}.
\newblock


\bibitem[Brase(2009)]%
        {brase2009datacite}
\bibfield{author}{\bibinfo{person}{Jan Brase}.}
  \bibinfo{year}{2009}\natexlab{}.
\newblock \showarticletitle{DataCite-A global registration agency for research
  data}. In \bibinfo{booktitle}{\emph{2009 fourth international conference on
  cooperation and promotion of information resources in science and
  technology}}. \bibinfo{publisher}{IEEE},
  \bibinfo{address}{\url{http://datacite.org}}, \bibinfo{pages}{257--261}.
\newblock


\bibitem[Breidenbach et~al\mbox{.}(2021)]%
        {chainlink}
\bibfield{author}{\bibinfo{person}{Lorenz Breidenbach},
  \bibinfo{person}{Christian Cachin}, \bibinfo{person}{Benedict Chan},
  \bibinfo{person}{Alex Coventry}, \bibinfo{person}{Steve Ellis},
  \bibinfo{person}{Ari Juels}, \bibinfo{person}{Farinaz Koushanfar},
  \bibinfo{person}{Andrew Miller}, \bibinfo{person}{Brendan Magauran},
  \bibinfo{person}{Daniel Moroz}, {et~al\mbox{.}}}
  \bibinfo{year}{2021}\natexlab{}.
\newblock \bibinfo{title}{Chainlink 2.0: Next steps in the evolution of
  decentralized oracle networks}.
\newblock \bibinfo{howpublished}{Chainlink Labs,
  \url{https://research.chain.link/whitepaper-v2.pdf}}.
\newblock


\bibitem[Buterin(2016)]%
        {buterin2016ethereum}
\bibfield{author}{\bibinfo{person}{Vitalik Buterin}.}
  \bibinfo{year}{2016}\natexlab{}.
\newblock \bibinfo{title}{What is ethereum?}
\newblock \bibinfo{howpublished}{Ethereum Official webpage. Available:
  \url{http://www.ethdocs.org/en/latest/introduction/what-is-ethereum.html}}.
\newblock


\bibitem[Chainlink(2022)]%
        {oracle2022}
\bibfield{author}{\bibinfo{person}{Chainlink}.}
  \bibinfo{year}{2022}\natexlab{}.
\newblock \bibinfo{title}{What is the blockchain oracle problem? why can't
  Blockchains Solve it?}
\newblock
\newblock
\urldef\tempurl%
\url{https://blog.chain.link/what-is-the-blockchain-oracle-problem/}
\showURL{%
\tempurl}


\bibitem[Connor(2011)]%
        {connor2011google}
\bibfield{author}{\bibinfo{person}{James Connor}.}
  \bibinfo{year}{2011}\natexlab{}.
\newblock \bibinfo{title}{Google Scholar citations open to all}.
\newblock \bibinfo{howpublished}{{Google Scholar Blog},
  \url{http://googlescholar.blogspot.com/2011/11/google-scholar-citations-open-to-all.html}}.
\newblock


\bibitem[Costas and Bordons(2008)]%
        {costas_bordons_2008}
\bibfield{author}{\bibinfo{person}{Rodrigo Costas} {and}
  \bibinfo{person}{María Bordons}.} \bibinfo{year}{2008}\natexlab{}.
\newblock \showarticletitle{Is G-index better than H-index? an exploratory
  study at the individual level}.
\newblock \bibinfo{journal}{\emph{Scientometrics}} \bibinfo{volume}{77},
  \bibinfo{number}{2} (\bibinfo{year}{2008}), \bibinfo{pages}{267–288}.
\newblock
\urldef\tempurl%
\url{https://doi.org/10.1007/s11192-007-1997-0}
\showDOI{\tempurl}


\bibitem[Egghe et~al\mbox{.}(2006)]%
        {gindex}
\bibfield{author}{\bibinfo{person}{Leo Egghe} {et~al\mbox{.}}}
  \bibinfo{year}{2006}\natexlab{}.
\newblock \showarticletitle{An improvement of the h-index: The g-index}.
\newblock \bibinfo{journal}{\emph{ISSI newsletter}} \bibinfo{volume}{2},
  \bibinfo{number}{1} (\bibinfo{year}{2006}), \bibinfo{pages}{8--9}.
\newblock


\bibitem[Fernandez(2010)]%
        {fernandez2010barriers}
\bibfield{author}{\bibinfo{person}{R Fernandez}.}
  \bibinfo{year}{2010}\natexlab{}.
\newblock \bibinfo{booktitle}{\emph{{Barriers to open science: from big
  business to Watson and Crick}}}.
\newblock {opensource.com supported by RedHat}.
\newblock
\urldef\tempurl%
\url{https://opensource.com/business/10/8/barriers-open-science-big-business-watson-and-crick}
\showURL{%
Retrieved Feb 06, 2021 from \tempurl}


\bibitem[Haak et~al\mbox{.}(2012)]%
        {haak2012orcid}
\bibfield{author}{\bibinfo{person}{Laurel~L Haak}, \bibinfo{person}{Martin
  Fenner}, \bibinfo{person}{Laura Paglione}, \bibinfo{person}{Ed Pentz}, {and}
  \bibinfo{person}{Howard Ratner}.} \bibinfo{year}{2012}\natexlab{}.
\newblock \showarticletitle{ORCID: a system to uniquely identify researchers}.
\newblock \bibinfo{journal}{\emph{Learned publishing}} \bibinfo{volume}{25},
  \bibinfo{number}{4} (\bibinfo{year}{2012}), \bibinfo{pages}{259--264}.
\newblock


\bibitem[Hirsch(2005)]%
        {pnas}
\bibfield{author}{\bibinfo{person}{Jorge~E Hirsch}.}
  \bibinfo{year}{2005}\natexlab{}.
\newblock \showarticletitle{An index to quantify an individual's scientific
  research output}.
\newblock \bibinfo{journal}{\emph{Proceedings of the National academy of
  Sciences}} \bibinfo{volume}{102}, \bibinfo{number}{46}
  (\bibinfo{year}{2005}), \bibinfo{pages}{16569--16572}.
\newblock


\bibitem[Hood and Sutherland(2021)]%
        {hood2021data}
\bibfield{author}{\bibinfo{person}{Amelia~SC Hood} {and}
  \bibinfo{person}{William~J Sutherland}.} \bibinfo{year}{2021}\natexlab{}.
\newblock \showarticletitle{The data-index: An author-level metric that values
  impactful data and incentivizes data sharing}.
\newblock \bibinfo{journal}{\emph{Ecology and Evolution}} \bibinfo{volume}{11},
  \bibinfo{number}{21} (\bibinfo{year}{2021}), \bibinfo{pages}{14344--14350}.
\newblock


\bibitem[Kelly and Jennions(2006)]%
        {kelly2006h}
\bibfield{author}{\bibinfo{person}{Clint~D Kelly} {and}
  \bibinfo{person}{Michael~D Jennions}.} \bibinfo{year}{2006}\natexlab{}.
\newblock \showarticletitle{The h index and career assessment by numbers}.
\newblock \bibinfo{journal}{\emph{Trends in Ecology \& Evolution}}
  \bibinfo{volume}{21}, \bibinfo{number}{4} (\bibinfo{year}{2006}),
  \bibinfo{pages}{167--170}.
\newblock


\bibitem[Kidwell et~al\mbox{.}(2016)]%
        {kidwell2016badges}
\bibfield{author}{\bibinfo{person}{Mallory~C Kidwell},
  \bibinfo{person}{Ljiljana~B Lazarevi{\'c}}, \bibinfo{person}{Erica Baranski},
  \bibinfo{person}{Tom~E Hardwicke}, \bibinfo{person}{Sarah Piechowski},
  \bibinfo{person}{Lina-Sophia Falkenberg}, \bibinfo{person}{Curtis Kennett},
  \bibinfo{person}{Agnieszka Slowik}, \bibinfo{person}{Carina Sonnleitner},
  \bibinfo{person}{Chelsey Hess-Holden}, {et~al\mbox{.}}}
  \bibinfo{year}{2016}\natexlab{}.
\newblock \showarticletitle{Badges to acknowledge open practices: A simple,
  low-cost, effective method for increasing transparency}.
\newblock \bibinfo{journal}{\emph{PLoS biology}} \bibinfo{volume}{14},
  \bibinfo{number}{5} (\bibinfo{year}{2016}), \bibinfo{pages}{e1002456}.
\newblock


\bibitem[King(2007)]%
        {king2007introduction}
\bibfield{author}{\bibinfo{person}{Gary King}.}
  \bibinfo{year}{2007}\natexlab{}.
\newblock \bibinfo{title}{An introduction to the dataverse network as an
  infrastructure for data sharing}.
\newblock , \bibinfo{numpages}{173--199}~pages.
\newblock


\bibitem[Ko and Park(2013)]%
        {ko2013index}
\bibfield{author}{\bibinfo{person}{Young~Man Ko} {and}
  \bibinfo{person}{Ji~Young Park}.} \bibinfo{year}{2013}\natexlab{}.
\newblock \showarticletitle{An index for evaluating journals in a small
  domestic citation index database whose citation rate is generally very low: A
  test based on the Korea Citation Index (KCI) database}.
\newblock \bibinfo{journal}{\emph{Journal of Informetrics}}
  \bibinfo{volume}{7}, \bibinfo{number}{2} (\bibinfo{year}{2013}),
  \bibinfo{pages}{404--411}.
\newblock


\bibitem[Koltun and Hafner(2021)]%
        {hbad}
\bibfield{author}{\bibinfo{person}{Vladlen Koltun} {and} \bibinfo{person}{David
  Hafner}.} \bibinfo{year}{2021}\natexlab{}.
\newblock \showarticletitle{The h-index is no longer an effective correlate of
  scientific reputation}.
\newblock \bibinfo{journal}{\emph{PloS one}} \bibinfo{volume}{16},
  \bibinfo{number}{6} (\bibinfo{year}{2021}), \bibinfo{pages}{e0253397}.
\newblock


\bibitem[McConaghy(2022)]%
        {ocean}
\bibfield{author}{\bibinfo{person}{Trent McConaghy}.}
  \bibinfo{year}{2022}\natexlab{}.
\newblock \bibinfo{title}{How Ocean Can Benefit Data Scientists}.
\newblock
  \bibinfo{howpublished}{\url{https://blog.oceanprotocol.com/how-ocean-can-benefit-data-scientists-7e502e5f1a5f}}.
\newblock


\bibitem[Morris et~al\mbox{.}(2011)]%
        {morris2011answer}
\bibfield{author}{\bibinfo{person}{Zo{\"e}~Slote Morris},
  \bibinfo{person}{Steven Wooding}, {and} \bibinfo{person}{Jonathan Grant}.}
  \bibinfo{year}{2011}\natexlab{}.
\newblock \showarticletitle{The answer is 17 years, what is the question:
  understanding time lags in translational research}.
\newblock \bibinfo{journal}{\emph{Journal of the Royal Society of Medicine}}
  \bibinfo{volume}{104}, \bibinfo{number}{12} (\bibinfo{year}{2011}),
  \bibinfo{pages}{510--520}.
\newblock


\bibitem[Olfson et~al\mbox{.}(2017)]%
        {olfson2017incentivizing}
\bibfield{author}{\bibinfo{person}{Mark Olfson}, \bibinfo{person}{Melanie~M
  Wall}, {and} \bibinfo{person}{Carlos Blanco}.}
  \bibinfo{year}{2017}\natexlab{}.
\newblock \showarticletitle{Incentivizing data sharing and collaboration in
  medical research—the s-index}.
\newblock \bibinfo{journal}{\emph{JAMA psychiatry}} \bibinfo{volume}{74},
  \bibinfo{number}{1} (\bibinfo{year}{2017}), \bibinfo{pages}{5--6}.
\newblock


\bibitem[Park and Wolfram(2019)]%
        {park2019research}
\bibfield{author}{\bibinfo{person}{Hyoungjoo Park} {and}
  \bibinfo{person}{Dietmar Wolfram}.} \bibinfo{year}{2019}\natexlab{}.
\newblock \showarticletitle{Research software citation in the Data Citation
  Index: Current practices and implications for research software sharing and
  reuse}.
\newblock \bibinfo{journal}{\emph{Journal of Informetrics}}
  \bibinfo{volume}{13}, \bibinfo{number}{2} (\bibinfo{year}{2019}),
  \bibinfo{pages}{574--582}.
\newblock


\bibitem[Pimm(2013)]%
        {unfairplayingfield}
\bibfield{author}{\bibinfo{person}{Jonathan Pimm}.}
  \bibinfo{year}{2013}\natexlab{}.
\newblock \showarticletitle{Scientific publishing - An unfair playing field}.
\newblock \bibinfo{journal}{\emph{The Psychiatrist}}  \bibinfo{volume}{37}
  (\bibinfo{date}{09} \bibinfo{year}{2013}), \bibinfo{pages}{281--282}.
\newblock
\urldef\tempurl%
\url{https://doi.org/10.1192/pb.bp.113.044768}
\showDOI{\tempurl}


\bibitem[Purkayastha et~al\mbox{.}(2019)]%
        {purkayastha2019comparison}
\bibfield{author}{\bibinfo{person}{Amrita Purkayastha},
  \bibinfo{person}{Eleonora Palmaro}, \bibinfo{person}{Holly~J
  Falk-Krzesinski}, {and} \bibinfo{person}{Jeroen Baas}.}
  \bibinfo{year}{2019}\natexlab{}.
\newblock \showarticletitle{Comparison of two article-level, field-independent
  citation metrics: Field-Weighted Citation Impact (FWCI) and Relative Citation
  Ratio (RCR)}.
\newblock \bibinfo{journal}{\emph{Journal of Informetrics}}
  \bibinfo{volume}{13}, \bibinfo{number}{2} (\bibinfo{year}{2019}),
  \bibinfo{pages}{635--642}.
\newblock


\bibitem[Reed et~al\mbox{.}(2020)]%
        {reed2020decentralized}
\bibfield{author}{\bibinfo{person}{Drummond Reed}, \bibinfo{person}{Manu
  Sporny}, \bibinfo{person}{Dave Longley}, \bibinfo{person}{Christopher Allen},
  \bibinfo{person}{Ryan Grant}, \bibinfo{person}{Markus Sabadello}, {and}
  \bibinfo{person}{Jonathan Holt}.} \bibinfo{year}{2020}\natexlab{}.
\newblock \bibinfo{booktitle}{\emph{{Decentralized Identifiers (DIDs) v1.0}}}.
\newblock {World Wide Web Consortium (W3C) Draft Community Group Report}.
\newblock


\bibitem[Rowhani-Farid et~al\mbox{.}(2017)]%
        {rowhani2017incentives}
\bibfield{author}{\bibinfo{person}{Anisa Rowhani-Farid},
  \bibinfo{person}{Michelle Allen}, {and} \bibinfo{person}{Adrian~G Barnett}.}
  \bibinfo{year}{2017}\natexlab{}.
\newblock \showarticletitle{What incentives increase data sharing in health and
  medical research? A systematic review}.
\newblock \bibinfo{journal}{\emph{Research integrity and peer review}}
  \bibinfo{volume}{2}, \bibinfo{number}{1} (\bibinfo{year}{2017}),
  \bibinfo{pages}{1--10}.
\newblock


\bibitem[Singh and Shetty(2017)]%
        {datasharing}
\bibfield{author}{\bibinfo{person}{Kritarth Naman~M. Singh} {and}
  \bibinfo{person}{Yashashri~C. Shetty}.} \bibinfo{year}{2017}\natexlab{}.
\newblock \showarticletitle{Data sharing: A viable resource for future}.
\newblock \bibinfo{journal}{\emph{Perspectives in clinical research}}
  \bibinfo{volume}{8,2} (\bibinfo{date}{apr--jun} \bibinfo{year}{2017}),
  \bibinfo{pages}{63--67}.
\newblock
\urldef\tempurl%
\url{https://doi.org/10.4103/2229-3485.203036}
\showDOI{\tempurl}


\bibitem[Sinha et~al\mbox{.}(2015)]%
        {sinha2015an}
\bibfield{author}{\bibinfo{person}{Arnab Sinha}, \bibinfo{person}{Zhihong
  Shen}, \bibinfo{person}{Yang Song}, \bibinfo{person}{Hao Ma},
  \bibinfo{person}{Darrin Eide}, \bibinfo{person}{Bo-June Hsu}, {and}
  \bibinfo{person}{Kuansan Wang}.} \bibinfo{year}{2015}\natexlab{}.
\newblock \showarticletitle{An overview of microsoft academic service (mas) and
  applications}. In \bibinfo{booktitle}{\emph{Proceedings of the 24th
  international conference on world wide web}}. \bibinfo{publisher}{ACM},
  \bibinfo{address}{\url{https://doi.org/10.1145/2740908.2742839}},
  \bibinfo{pages}{243--246}.
\newblock


\bibitem[Sugiura et~al\mbox{.}(2012)]%
        {drugs}
\bibfield{author}{\bibinfo{person}{Lisa Sugiura}, \bibinfo{person}{Catherine
  Pope}, {and} \bibinfo{person}{Craig Webber}.}
  \bibinfo{year}{2012}\natexlab{}.
\newblock \showarticletitle{Buying unlicensed slimming drugs from the Web: a
  virtual ethnography}. In \bibinfo{booktitle}{\emph{Proceedings of the 4th
  Annual ACM Web Science Conference}}. \bibinfo{publisher}{{ACM}},
  \bibinfo{address}{\url{https://doi.org/10.1145/2380718.2380755}},
  \bibinfo{pages}{284--287}.
\newblock


\bibitem[Valenzuela et~al\mbox{.}(2015)]%
        {meaningfulcitations}
\bibfield{author}{\bibinfo{person}{Marco Valenzuela}, \bibinfo{person}{Vu Ha},
  {and} \bibinfo{person}{Oren Etzioni}.} \bibinfo{year}{2015}\natexlab{}.
\newblock \showarticletitle{Identifying Meaningful Citations.}. In
  \bibinfo{booktitle}{\emph{AAAI workshop: Scholarly big data}},
  Vol.~\bibinfo{volume}{15}. \bibinfo{publisher}{Workshops at the twenty-ninth
  AAAI conference on artificial intelligence},
  \bibinfo{address}{\url{https://ai2-website.s3.amazonaws.com/publications/ValenzuelaHaMeaningfulCitations.pdf}},
  \bibinfo{pages}{13}.
\newblock


\end{thebibliography}
\end{document}